\documentclass[11pt,a4paper]{article}

\usepackage[margin=2.2cm]{geometry}
\usepackage{amsmath}
\usepackage{amssymb}
\usepackage{graphicx}
\usepackage{booktabs}
\usepackage{caption}
\usepackage{hyperref}

\title{On the alleged chaos in periodically forced traveling-wave reductions of fractional WBBM models: a quantitative re-examination}
\author{Kaixuan Niu\thanks{Corresponding author: niukaixuan@ginchin.cn}\\China Science Trans-century Aerospace Institute}
\date{}

\begin{document}

\maketitle

\begin{abstract}
A large and growing literature applies a standard pipeline to fractional nonlinear evolution equations --- traveling-wave reduction to a planar Hamiltonian system, addition of a periodic forcing term, and visual inspection of phase portraits --- to claim bifurcations, quasi-periodicity, and chaos. In much of this literature the chaos assertions rest exclusively on the visual appearance of phase diagrams and time series, without Lyapunov exponents, Poincar\'e sections, or any other quantitative diagnostic. This practice is particularly problematic because the reduced systems are typically undamped, near-integrable oscillators, for which KAM theory predicts that most orbits lie on invariant tori and chaos, if present at all, is confined to thin stochastic layers: the burden of proof for a chaos claim in such systems is correspondingly high. In this paper we articulate the methodological pitfalls of the pipeline and subject a recent representative example --- Ullah, Ali and Roshid's study of the second fractional Wazwaz--Benjamin--Bona--Mahony (WBBM) model [PLoS ONE 19(7): e0307565 (2024)] --- to a full quantitative re-examination, combining analytical arguments with four independent numerical diagnostics: Benettin largest Lyapunov exponents in two cross-validated configurations, an unrenormalized separation-growth test over $3\times10^{6}$ time units, stroboscopic Poincar\'e sections, and spectral analysis. For the example paper we find that (i) its linear stability analysis concludes ``unstable propagation'' from a dispersion relation that is real for every real wave number --- all modes are in fact neutrally stable, and the reported singularity is a pole (resonant surface), not a temporal instability; (ii) its unforced system, labeled ``quasi-periodic'', is necessarily periodic, being a two-dimensional autonomous Hamiltonian system; (iii) all four of its chaos assertions fail every quantitative diagnostic --- the largest Lyapunov exponents are bounded by $5\times10^{-7}$, separation growth is linear in time, and the Poincar\'e sections are smooth closed invariant curves --- while its single quasi-periodic assertion is confirmed; and (iv) its equilibrium classification and phase portraits are correct. We further document a false positive of the Gottwald--Melbourne 0--1 test ($K \approx 0.80$) on one of these regular orbits, caused by dense spectral line clusters from a slow modulation, and show that stroboscopic parameter sweeps of such conservative systems produce broadband, ``chaotic-looking'' diagrams for purely regular tori. We close with an actionable checklist of quantitative standards for chaos claims in this literature.
\end{abstract}


\section{Introduction}
\label{sec:intro}

\subsection{The pipeline and its evidentiary gap}
\label{sec:pipeline}

Over the past decade a recognizable template has become widespread in the nonlinear-wave literature, particularly in studies of fractional-order evolution equations: a partial differential equation is reduced by a traveling-wave ansatz to an ordinary differential equation; the ODE is recast as a planar Hamiltonian system, whose equilibria and phase portraits are classified by the planar-dynamical method of Liu and Li \cite{liu2002}; a periodic forcing term is then added, and the resulting non-autonomous oscillator is integrated numerically for a handful of parameter values; finally, the appearance of the 2D/3D phase portraits and time series is used to pronounce the dynamics ``periodic'', ``quasi-periodic'', or ``chaotic''. Explicit soliton solutions are read off from the Hamiltonian level curves, and a linear dispersion analysis is often appended. Dozens of papers following this template appear every year; the reference lists of any one of them typically contain many others.

The analytical parts of this pipeline --- equilibrium classification, Hamiltonian structure, exact solutions --- are usually executed correctly. The weak point is the dynamical classification of the forced system: with very few exceptions, the labels ``chaotic'' and ``quasi-periodic'' are assigned by visual inspection alone, and quantitative diagnostics --- Lyapunov exponents with demonstrated convergence, stroboscopic Poincar\'e sections, bifurcation diagrams with identified periodic windows, spectral analysis --- are almost never reported. This stands in contrast to long-articulated community standards: Sprott's proposed standard for the publication of new chaotic systems \cite{sprott2011} asks explicitly for quantitative evidence of sensitive dependence, and it is telling that papers in this genre routinely cite that standard (as the example studied below does) without applying it.

\subsection{Why visual inspection fails precisely for these systems}
\label{sec:visual}

The evidentiary gap matters more here than in a generic dissipative setting, because of the structure of the reduced systems: the traveling-wave reduction of a conservative dispersive PDE yields a planar Hamiltonian system, and the ``perturbed'' system studied for chaos is that system plus a periodic forcing, with no damping --- a one-degree-of-freedom, undamped, periodically forced Hamiltonian oscillator. For such systems the Kolmogorov--Arnold--Moser (KAM) theory \cite{lichtenberg1992} provides a rigid prior: most invariant tori of the integrable limit survive small perturbation, generic initial conditions lie on quasi-periodic tori, and chaos --- where it exists at all --- is confined to thin stochastic layers around resonances and separatrices; a \emph{single-well} oscillator has not even a separatrix. Two consequences follow. First, the a priori probability that a randomly chosen parameter set and initial condition lands on a chaotic orbit is small, so chaos claims demand positive quantitative proof rather than suggestive pictures. Second, the visual distinction between a quasi-periodic torus and a thin stochastic layer is subtle: stroboscopic projections of tori densely fill bands (demonstrated explicitly in Section~\ref{sec:morphology}), and multi-frequency quasi-periodic time series look irregular to the untrained eye. A phase portrait in a conservative system cannot, by itself, separate order from chaos.

A related pitfall concerns the statistical tests that are sometimes proposed as quick fixes. The Gottwald--Melbourne 0--1 test \cite{gottwald2004} is attractive because it requires only a scalar time series and no phase-space reconstruction, and it performs well on the dissipative examples for which it was designed. Its behavior on conservative, slowly modulated orbits is less benign: we document below a case in which the test returns a median $K = 0.80$ --- a clear ``chaotic'' verdict by its own criterion --- for an orbit that four independent diagnostics prove to be a regular torus. The mechanism (a slow modulation producing dense spectral line clusters that inflate the mean square displacement within any practical truncation window) is structural, not an implementation artifact. To our knowledge this failure mode has not been quantitatively documented in the context of the traveling-wave-reduction literature, and we regard its demonstration as one of the contributions of this paper.

\subsection{A representative case study}
\label{sec:casestudy}

Rather than criticizing the genre in the abstract, we selected a recent, typical, and fully specified instance and re-examined every one of its dynamical claims with analytical and numerical care: the study by Ullah, Ali and Roshid of the second fractional Wazwaz--Benjamin--Bona--Mahony (WBBM) model \cite{ullah2024} (henceforth ``the example paper''). The example paper is a fair representative of the pipeline: it performs the traveling-wave reduction of the second fractional 3D WBBM equation to a planar Hamiltonian system, classifies four phase-portrait regimes, adds a forcing $\sigma\cos(\omega\mathcal{B})$ and asserts quasi-periodic behavior at $\sigma = 0$ and chaos at four further parameter/initial-condition combinations, derives elliptic-function and soliton solutions, and appends a linear stability analysis concluding ``unstable propagation'' for certain wave numbers. All parameter values are stated, which makes an exact re-examination possible.

Our findings for the example paper, developed in Sections~\ref{sec:case1}--\ref{sec:case3}, are: (i) the linear stability conclusion is incorrect --- the dispersion relation is real for all real wave numbers, every Fourier mode is neutrally stable, and the reported ``unstable propagation'' is a pole of the dispersion relation misread as a temporal instability; (ii) the unforced regime labeled ``quasi-periodic'' is necessarily periodic, as a matter of Hamiltonian structure, confirmed by spectral analysis at energy conservation of $2.7\times10^{-15}$; (iii) none of the four chaos assertions survives quantitative testing --- largest Lyapunov exponents are bounded by $5\times10^{-7}$, separation growth is linear in time with log--log slopes $0.96$--$1.01$ over $3\times10^{6}$ time units, and all Poincar\'e sections are smooth closed invariant curves --- while the paper's single quasi-periodic assertion is confirmed; and (iv) the equilibrium classification and phase portraits of the bifurcation analysis are correct and were independently reproduced. Section~\ref{sec:remarks} collects secondary issues (including the triviality of the conformable-derivative framework under $\tau = t^{\gamma}/\gamma$), Section~\ref{sec:recommendations} formulates an actionable checklist of quantitative standards for chaos claims in this literature, Section~\ref{sec:conclusions} concludes, and an Appendix details all numerical methods.

\section{The example paper and its reduction}
\label{sec:example}

The example paper \cite{ullah2024} studies the second fractional three-dimensional WBBM equation,
\begin{equation*}
D_t^{\gamma} w + D_z^{\gamma} w + D_x^{\gamma}\!\left(w^3\right) - D_{xyt}^{3\gamma} w = 0,
\end{equation*}
where $D^{\gamma}$ denotes the conformable derivative of order $\gamma \in (0,1]$ \cite{khalil2014}. The traveling-wave reduction $\mathcal{B} = \frac{1}{\gamma}\left(l_1 x^{\gamma} + l_2 y^{\gamma} + l_3 z^{\gamma} - l_4 t^{\gamma}\right)$ leads, after one integration, to the planar dynamical system
\begin{equation}
\frac{dW}{d\mathcal{B}} = P, \qquad \frac{dP}{d\mathcal{B}} = -\alpha W^3 - mW,
\label{eq:planar}
\end{equation}
with $\alpha = 1/(l_2 l_4)$ and $m = (l_3 - l_4)/(l_1 l_2 l_4)$, possessing the Hamiltonian
\begin{equation}
H(W,P) = \frac{1}{2}P^2 + \frac{\alpha}{4}W^4 + \frac{m}{2}W^2.
\label{eq:hamiltonian}
\end{equation}

The example paper then (i) classifies equilibria and phase portraits in four parameter regimes (its Fig.~1); (ii) adds a periodic forcing,
\begin{equation}
\frac{dW}{d\mathcal{B}} = P, \qquad \frac{dP}{d\mathcal{B}} = -\alpha W^3 - mW + \sigma\cos(\omega\mathcal{B}),
\label{eq:forced}
\end{equation}
with $\alpha = 3$, $m = 1$ (a single-well hardening Duffing oscillator), and asserts quasi-periodic behavior for $\sigma = 0$ (its Fig.~2), chaos for $(\sigma, \omega) = (0.3, 2.2)$ and $(1.4, 3.9)$ (its Figs.~3--4), and, for $(\sigma, \omega) = (1.9, 3.9)$, quasi-periodicity from the initial condition $(0.8, 0)$ and chaos from $(0, 0.01)$ and $(0.1, 0.2)$ (its Fig.~5); (iii) constructs Jacobi elliptic, kink and soliton solutions from \eqref{eq:hamiltonian}; and (iv) performs a linear stability analysis concluding ``stable propagation'' for some wave numbers and ``unstable propagation'' for $b = 1$, $d = 1$, $c \to -1$ (its Fig.~10).

\noindent\textbf{What survives scrutiny.} Before turning to the claims that fail, we record what withstands verification. We independently reconstructed the phase portraits of system \eqref{eq:planar} as level curves of the Hamiltonian \eqref{eq:hamiltonian}, classifying equilibria analytically through $\lambda^2 = -3\alpha W_0^2 - m$ at each equilibrium $(W_0, 0)$. For the four parameter regimes of the example paper we obtain:

\begin{table}[htbp]
\centering
\small
\begin{tabular}{@{}clcclc@{}}
\toprule
Case & Parameters & $\alpha$ & $m$ & Equilibria & Agreement \\
\midrule
1 & $l_4 = 2$, $l_1 = l_3 = 1$, $l_2 = -1$ & $-1/2$ & $+1/2$ & $(0,0)$ center; $(\pm 1, 0)$ saddles & yes \\
2 & $l_4 = 2$, $l_1 = l_2 = l_3 = 1$ & $+1/2$ & $-1/2$ & $(0,0)$ saddle; $(\pm 1, 0)$ centers & yes \\
3 & $l_1 = l_3 = 1$, $l_2 = l_4 = -1$ & $+1$ & $+2$ & $(0,0)$ center & yes \\
4 & $l_1 = l_2 = l_3 = 1$, $l_4 = -1$ & $-1$ & $-2$ & $(0,0)$ saddle & yes \\
\bottomrule
\end{tabular}
\end{table}

\begin{figure}[htbp]
\centering
\includegraphics[width=0.85\textwidth]{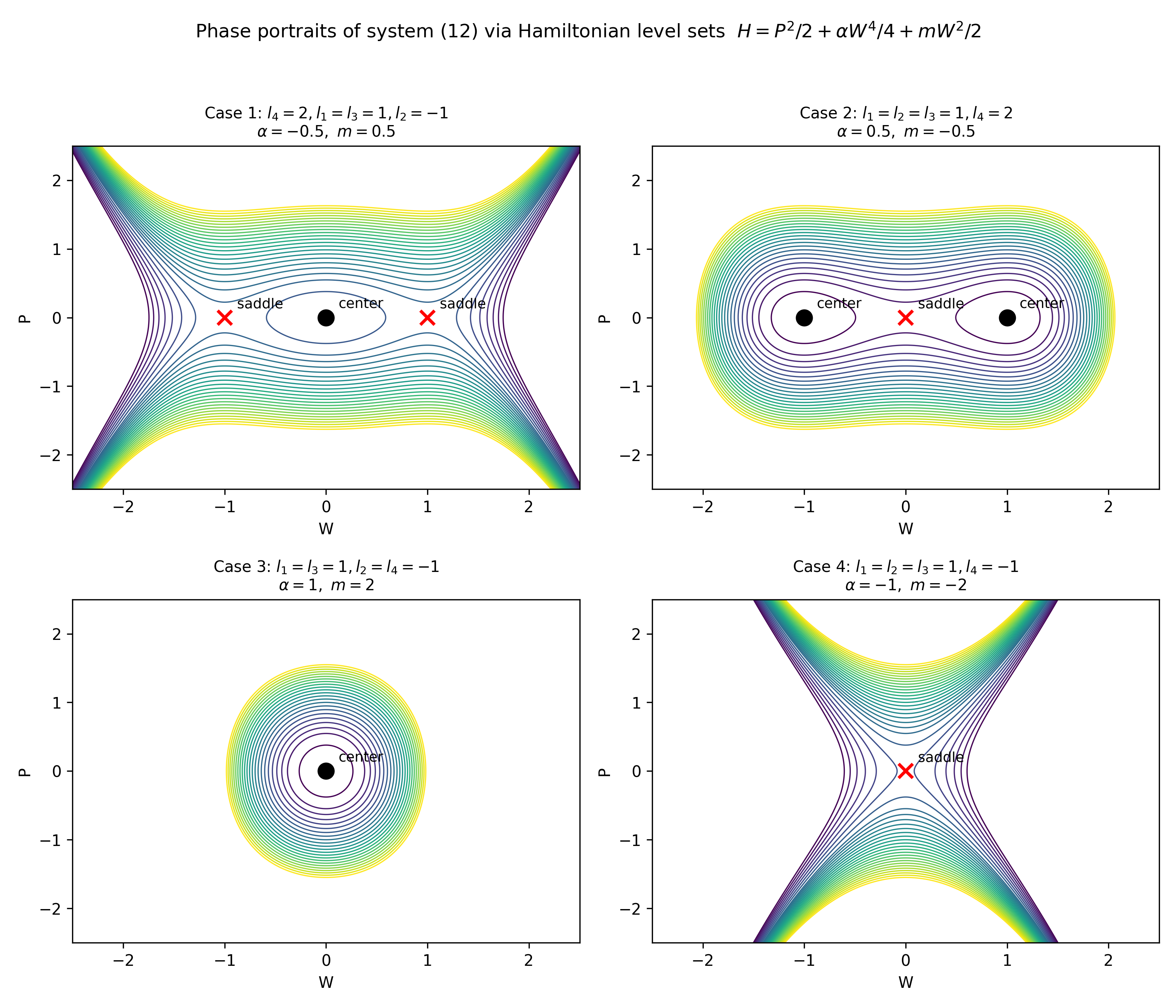}
\caption{Independent reconstruction of the phase portraits of system \eqref{eq:planar} as Hamiltonian level curves $H(W,P) = h$ for the four parameter regimes of the example paper: (a) $\alpha = -1/2$, $m = 1/2$; (b) $\alpha = 1/2$, $m = -1/2$; (c) $\alpha = 1$, $m = 2$; (d) $\alpha = -1$, $m = -2$. Equilibrium types (centers, saddles) agree in all four cases with the example paper's Fig.~1.}
\label{fig:phase}
\end{figure}

The reconstructed phase portraits (Fig.~\ref{fig:phase}) coincide with the example paper's Fig.~1, including the heteroclinic connections of Case 1 and the homoclinic figure-eight of Case 2. The bifurcation/equilibrium analysis, executed with the standard method of Liu and Li \cite{liu2002}, is correct. In addition, the quasi-periodic classification of one forced orbit (Case D1 in Section~\ref{sec:case3}) is confirmed by our diagnostics. The problems of the example paper are therefore not in its analytical core but in exactly those claims that rest on interpretation rather than computation --- which is where the pipeline genre as a whole is weakest.

\section{Case study I: the linear stability analysis cannot support ``unstable propagation''}
\label{sec:case1}

\noindent\textbf{The claim.} Section~9 of the example paper perturbs a constant state $a$ of the governing equation, linearizes, inserts the plane-wave ansatz
\begin{equation}
F(t,x,y,z) = e^{\,i(bx + cy + dz - \rho t)},
\label{eq:ansatz}
\end{equation}
and obtains the dispersion relation (its Eq.~(28))
\begin{equation}
\rho(b,c,d) = \frac{d + a^2 b}{1 + bc}, \qquad bc \neq -1.
\label{eq:dispersion}
\end{equation}
It then states that \eqref{eq:dispersion} ``remains stable propagation for $b = 1, c = 1$'' but ``remains unstable propagation for $b = 1, d = 1$, since for the selected parameters, the denominator $bc + 1$ approaches zero when $c$ is chosen close to $-1$'' (its Fig.~10).

\noindent\textbf{Refutation.} The derivation of \eqref{eq:dispersion} is itself correct; we verified symbolically that substituting \eqref{eq:ansatz} into the linearized equation
\begin{equation}
a^2 F_x + F_t + F_z - F_{xyt} = 0
\label{eq:linearized}
\end{equation}
and dividing by $F$ leaves the residual $i\left(a^2 b - bc\rho + d - \rho\right)$, whose vanishing yields exactly \eqref{eq:dispersion}. The error lies in the interpretation. For real wave numbers $a, b, c, d \in \mathbb{R}$ with $bc \neq -1$, both the numerator and the denominator of \eqref{eq:dispersion} are real, hence
\begin{equation*}
\rho(b,c,d) \in \mathbb{R} \quad \text{for all admissible } (b,c,d).
\end{equation*}
Consequently $|F(t,x,y,z)| = \left|e^{i(bx+cy+dz - \rho t)}\right| = 1$ for all $t$: no Fourier mode grows or decays in time. A modal (temporal) instability requires a mode with $\operatorname{Im}\rho > 0$; the dispersion relation \eqref{eq:dispersion} admits no such mode anywhere in its domain. The linearized system \eqref{eq:linearized} is therefore \textbf{neutrally stable}, for all parameter choices, including $b = 1$, $d = 1$, $c \to -1$.

\begin{figure}[htbp]
\centering
\includegraphics[width=0.85\textwidth]{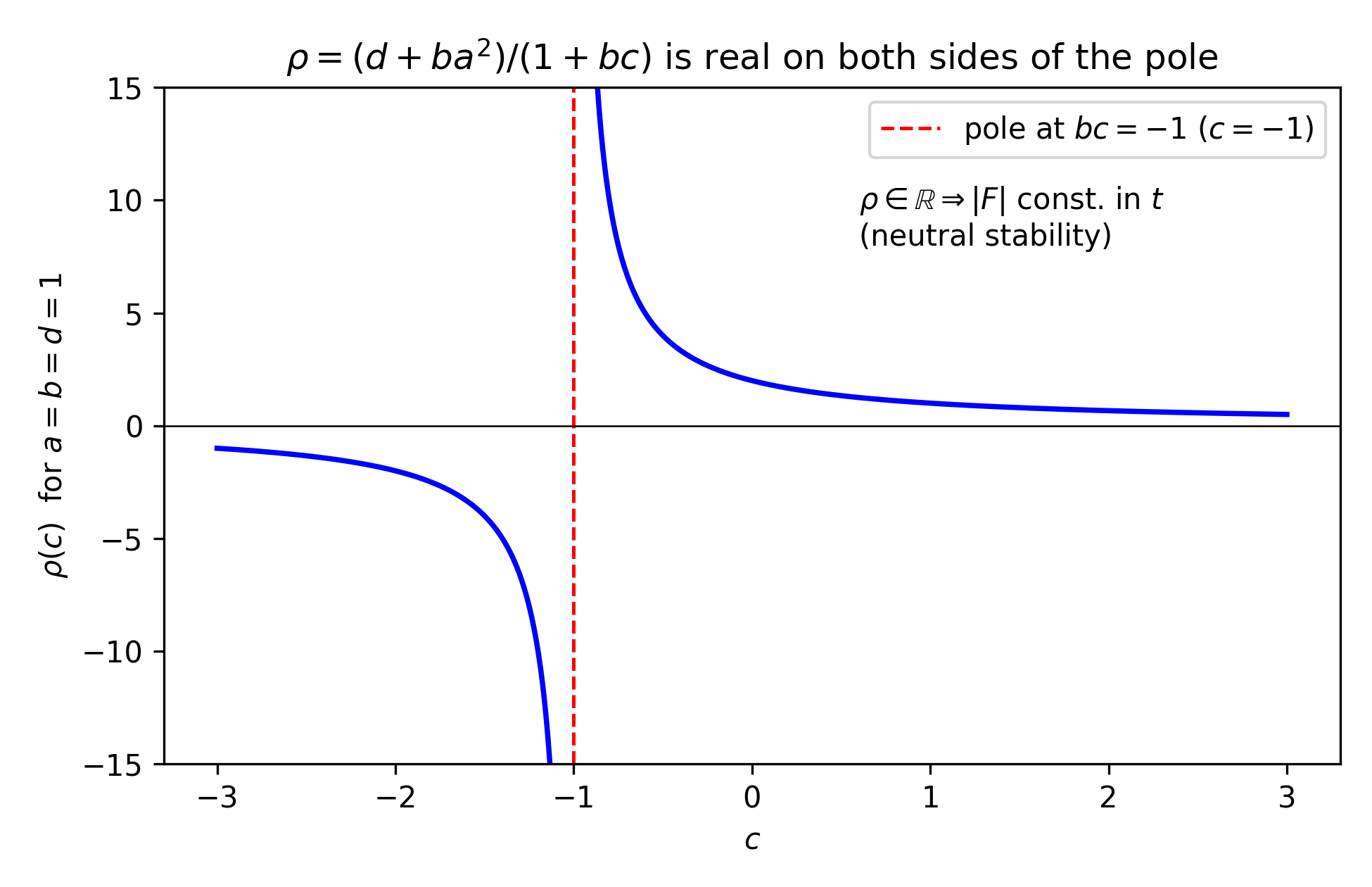}
\caption{The dispersion relation \eqref{eq:dispersion} for $a = b = d = 1$ as a function of $c$. The function is real on both sides of the pole at $c = -1$ (dashed line); $|F|$ is therefore constant in time for every admissible wave number, i.e.\ all modes are neutrally stable. The pole is a resonance of the ansatz \eqref{eq:ansatz}, not a dynamical instability, which would require $\operatorname{Im}\rho > 0$.}
\label{fig:stability}
\end{figure}

The set $bc = -1$ on which \eqref{eq:dispersion} diverges is a \emph{pole} of the dispersion relation --- a resonant surface on which the plane-wave ansatz \eqref{eq:ansatz} simply fails to produce a solution of \eqref{eq:linearized} (the compatibility condition $a^2 b + d = 0$ would be required for a nontrivial limit, and generically fails). A singularity of the function $\rho(b,c,d)$ in \emph{wave-number space} is not a growth of $F$ in \emph{time}. The example paper conflates ``the real function $\rho$ is unbounded near $c = -1$'' with ``the perturbation propagates unstably''. These are logically independent statements; the first is true and the second is false. Note in particular that even arbitrarily large $|\rho|$ only implies rapid oscillation, since the temporal factor is $e^{-i\rho t}$ with unit modulus. Figure~\ref{fig:stability} displays $\rho(c)$ for $a = b = d = 1$: the two real branches blow up at the pole $c = -1$ while remaining real everywhere, which is precisely the neutral, dispersive behavior described above.

\noindent\textbf{Correct statement.} The linear stability analysis of the example paper establishes that all Fourier modes of the linearized equation are neutrally stable (purely oscillatory), and that the plane-wave ansatz is incompatible with wave numbers on the resonant surface $bc = -1$. No conclusion about ``unstable propagation'' can be drawn from this analysis.

Two minor points complete the picture. First, the example text states that the analysis is applied to ``the integer order, as specified in equation Eq.~(4)'' of that paper; however, the perturbation equations actually used (its Eqs.~(24)--(25)) contain the derivative structure $(F_t, F_z, F_x, F_{xyt})$, which is that of its Eq.~(5) --- the second WBBM equation --- not of its Eq.~(4), whose dispersion term is $w_{xzt}$ and whose nonlinearity differentiates in $y$. Second, labeling a constant background $a$ a ``stable-state solution'' is assumed, not verified, in that analysis. The general lesson extends beyond the example: in this literature, dispersion-relation plots are frequently used as stability verdicts; the only quantity that decides temporal stability is the sign of $\operatorname{Im}\rho$, and a real-valued dispersion relation can never certify instability.

\section{Case study II: the $\sigma = 0$ regime is periodic, not quasi-periodic}
\label{sec:case2}

\noindent\textbf{The claim.} For $\sigma = 0$, initial condition $(0, 0.01)$, the example paper states that system \eqref{eq:forced} ``exhibits quasi-periodic behavior in time series plots and 2D and 3D phase diagrams'' (its Fig.~2).

\noindent\textbf{Refutation (theory).} For $\sigma = 0$, system \eqref{eq:forced} reduces to the autonomous planar Hamiltonian system \eqref{eq:planar} with Hamiltonian \eqref{eq:hamiltonian}. Every bounded orbit of a two-dimensional autonomous Hamiltonian system lies on a connected component of a level curve $H(W,P) = h$; for $\alpha > 0$, $m > 0$ every such component enclosing the single center $(0,0)$ is a simple closed curve, and the motion on it is periodic. Quasi-periodic motion requires at least two incommensurate frequencies, which is impossible in a one-degree-of-freedom autonomous Hamiltonian system. The ``quasi-periodic'' label is therefore excluded on structural grounds, before any numerics.

\begin{figure}[htbp]
\centering
\includegraphics[width=0.85\textwidth]{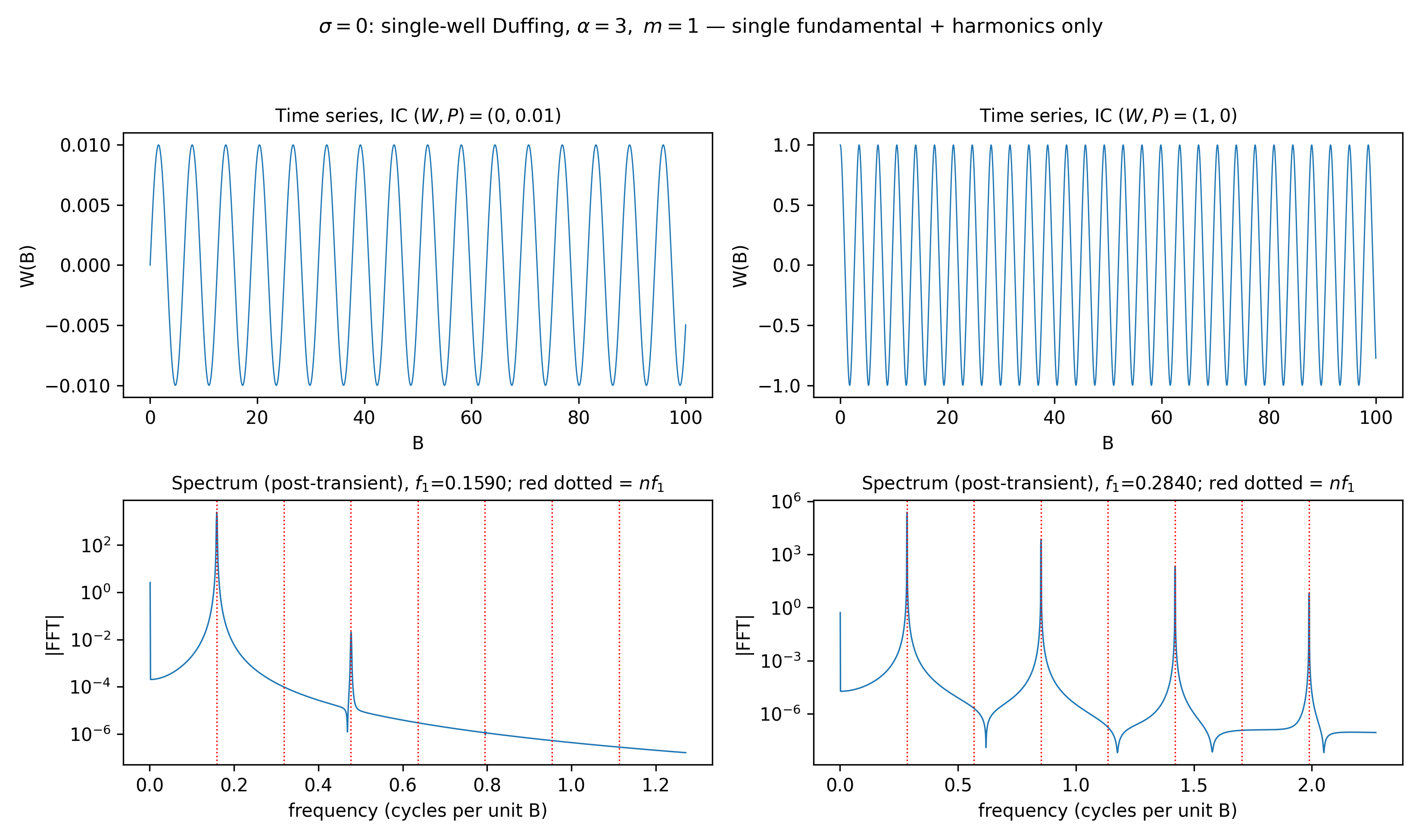}
\caption{FFT spectra of $W(\mathcal{B})$ for the unforced system ($\sigma = 0$), $\alpha = 3$, $m = 1$. (a) Initial condition $(0, 0.01)$: a single fundamental and its third harmonic; energy drift $|\Delta H| \leq 2.7\times10^{-15}$ over 2000 time units. (b) Initial condition $(1.0, 0)$: fundamental with odd harmonics $3f_1, 5f_1, 7f_1$. In both cases all peaks are commensurate, confirming periodic rather than quasi-periodic motion.}
\label{fig:sigma0}
\end{figure}

\noindent\textbf{Refutation (computation).} We integrated the $\sigma = 0$ system with a DOP853 integrator at tolerances $\mathrm{rtol} = 10^{-12}$, $\mathrm{atol} = 10^{-14}$ over $\mathcal{B} \in [0, 2000]$, sampled uniformly at $d\mathcal{B} = 10^{-3}$, and computed FFT spectra (Hann window, first half of the record discarded; only peaks with amplitude $\geq 10^{-6}$ of the dominant peak retained). For the initial condition $(0, 0.01)$ the energy is conserved to $\max|\Delta H| = 2.7\times10^{-15}$, and the spectrum contains exactly one fundamental $f_1 = 0.159134$ cycles/$\mathcal{B}$ (angular frequency $\approx 0.9990$, consistent with the linear frequency $\sqrt{m} = 1$ of the well) together with its third harmonic at relative amplitude $8.2\times10^{-6}$; the measured frequency ratios $1.0008$ and $3.0033$ match $1$ and $3$ to within Hann-window broadening. For the larger-amplitude initial condition $(1.0, 0)$ ($\max|\Delta H| = 4.4\times10^{-10}$), the spectrum contains the fundamental $f_1 = 0.284000$ and the odd harmonics $3f_1, 5f_1, 7f_1$ with commensurability to $10^{-4}$ accuracy and rapidly decreasing amplitudes ($3.0\times10^{-2}$, $8.9\times10^{-4}$, $2.6\times10^{-5}$) --- the textbook odd-harmonic signature of a periodic orbit in a symmetric potential. No second, incommensurate fundamental frequency exists in either case (Fig.~\ref{fig:sigma0}).

\noindent\textbf{Correct statement.} The orbit displayed in the example paper's Fig.~2 is a periodic orbit (a closed curve in the $(W,P)$ plane), not a quasi-periodic one. The apparent complexity of its 3D embedding and time series is a plotting artifact of the projection, not a dynamical phenomenon. The general lesson: in a pipeline whose unforced limit is a two-dimensional autonomous Hamiltonian system, ``quasi-periodic'' at zero forcing is not a possible classification, and its appearance in a paper is a reliable indicator that dynamical labels were assigned by inspection.

\section{Case study III: the chaos assertions are not supported by quantitative diagnostics}
\label{sec:case3}

This section contains our main computational result. We show that none of the four parameter sets labeled chaotic in the example paper exhibits sensitive dependence on initial conditions, and that all of them are regular orbits on invariant tori.

\subsection{Why quantitative diagnostics are indispensable here}
\label{sec:diagnostics}

System \eqref{eq:forced} with $\alpha = 3$, $m = 1$ is an \emph{undamped}, \emph{single-well}, periodically forced oscillator --- precisely the setting of Section~\ref{sec:visual}, in which KAM theory \cite{lichtenberg1992} makes quasi-periodic tori the generic outcome and confines any chaos to thin stochastic layers. Chaos assertions here therefore carry a substantial burden of proof. The example paper supports its chaos claims almost exclusively with 3D/2D phase portraits and time-series plots; the Poincar\'e plot and ``Lyapunov plot'' in its Fig.~5 appear without any statement of method (see below). As Section~\ref{sec:morphology} demonstrates explicitly, stroboscopic projections of quasi-periodic tori in conservative systems fill bands and produce pictures that ``look chaotic everywhere''; visual inspection cannot distinguish them from genuine chaos. We also note that the example paper's own citation [42] therein --- Sprott's proposed standard for the publication of new chaotic systems \cite{sprott2011} --- asks for quantitative evidence such as Lyapunov exponents and bifurcation diagrams, none of which is provided; the ``Lyapunov plot'' in the example paper's Fig.~5c appears without any statement of algorithm, initial separation, renormalization interval, or convergence.

\subsection{Methods}
\label{sec:methods}

We tested the five parameter/initial-condition sets summarized in Table~\ref{tab:diagnostics}: Case B (example paper's Fig.~3, $\sigma = 0.3$, $\omega = 2.2$), Case C (its Fig.~4, $\sigma = 1.4$, $\omega = 3.9$), and Cases D1, D2, D3 (its Fig.~5, $\sigma = 1.9$, $\omega = 3.9$, initial conditions $(0.8, 0)$, $(0, 0.01)$, $(0.1, 0.2)$ respectively). Four independent diagnostics were used; full details are given in the Appendix.

\noindent\textbf{(a) Benettin largest Lyapunov exponent} \cite{benettin1980}: two-trajectory algorithm with initial separation $d_0 = 10^{-9}$, total integration length $2\times10^{5}$ time units after a transient of $10^{4}$, in two cross-validating configurations (renormalization interval $\Delta\mathcal{B} = 1$, step $10^{-3}$; and $\Delta\mathcal{B} = 5$, step $5\times10^{-4}$); uncertainties are standard errors of the mean over 25 blocks.

\noindent\textbf{(b) Unrenormalized separation growth} (decisive diagnostic): two trajectories separated by $d_0 = 10^{-9}$ are integrated for $3\times10^{6}$ time units \emph{without} renormalization. For a chaotic orbit, $\log d(t)$ grows linearly in $t$ with slope $\lambda_{\max}$; for a regular orbit, separation grows linearly in time, i.e.\ $\log d$ grows linearly in $\log t$ with slope $\approx 1$. This diagnostic is immune to the saturation artifacts that can bias renormalized exponents either way and directly distinguishes exponential from polynomial growth.

\noindent\textbf{(c) Stroboscopic Poincar\'e sections}: points recorded once per forcing period, $\geq 3.5\times10^{4}$ periods after transient. Chaotic orbits produce area-filling point clouds; quasi-periodic orbits produce smooth closed invariant curves.

\noindent\textbf{(d) Spectral analysis} of long high-accuracy records (rtol $10^{-11}$): regular orbits have discrete line spectra; chaos has broadband continuous spectra.

Additionally, we ran the \textbf{Gottwald--Melbourne 0--1 test} \cite{gottwald2004} (100 random values $c \in (0,\pi)$, $N = 2\times10^{5}$, $n_{\mathrm{cut}} = N/10$, exact FFT-based mean square displacement with oscillatory term removed; implementation calibrated on synthetic periodic/quasi-periodic signals giving $K \approx 0$ and on the logistic map at $r = 3.9$ and white noise giving $K \approx 0.998$). The result is reported because it is instructive, not because we rely on it (Section~\ref{sec:01test}).

\subsection{Results}
\label{sec:results}

\begin{table}[htbp]
\centering
\small
\caption{Quantitative diagnostics for the five parameter sets of the example paper's chaos section (transposed layout; rows are diagnostics, columns are cases). All $\lambda$ values are Benettin largest exponents (main and robust configurations, 25-block SEM), in units of $10^{-5}$; ``exp.\ slope'' is the fitted slope of $\log d$ vs.\ $t$ and ``log slope'' the fitted slope of $\log d$ vs.\ $\log t$ from diagnostic (b) over $3\times10^{6}$ time units; $K$ is the median of the 0--1 test; the last row states our verdict on the example paper's assertion.}
\label{tab:diagnostics}
\begin{tabular}{@{}lccccc@{}}
\toprule
 & B (Fig.~3) & C (Fig.~4) & D1 (Fig.~5) & D2 (Fig.~5) & D3 (Fig.~5) \\
\midrule
$\sigma$ & 0.3 & 1.4 & 1.9 & 1.9 & 1.9 \\
$\omega$ & 2.2 & 3.9 & 3.9 & 3.9 & 3.9 \\
IC & $(0,0.01)$ & $(0,0.01)$ & $(0.8,0)$ & $(0,0.01)$ & $(0.1,0.2)$ \\
Claim in \cite{ullah2024} & chaotic & chaotic & quasi-periodic & chaotic & chaotic \\
$\lambda$ main ($10^{-5}$) & $3.42 \pm 1.8$ & $3.90 \pm 2.2$ & $4.27 \pm 4.2$ & $3.83 \pm 2.2$ & $5.36 \pm 3.6$ \\
$\lambda$ robust ($10^{-5}$) & $3.42 \pm 1.8$ & $3.75 \pm 2.2$ & $3.73 \pm 3.9$ & $4.06 \pm 2.2$ & $3.35 \pm 3.6$ \\
exp.\ slope & $4.6\times10^{-7}$ & $4.6\times10^{-7}$ & $2.7\times10^{-7}$ & $4.6\times10^{-7}$ & $4.4\times10^{-7}$ \\
log slope & 1.003 & 1.003 & 0.611 (saturating) & 1.008 & 0.964 \\
$K$ median & 0.381 & 0.800 & 0.223 & 0.097 & 0.681 \\
Poincar\'e section & \shortstack{regular band,\\island chain} & \shortstack{thin closed ring\\(2-torus)} & closed curve & closed curve & closed curve \\
Claim holds? & \textbf{No} & \textbf{No} & \textbf{Yes} & \textbf{No} & \textbf{No} \\
\bottomrule
\end{tabular}
\end{table}

Three observations emerge.

\noindent\emph{(i) The Benettin values are integration artifacts, not positive exponents.} In all five cases the running estimate of $\lambda$ decreases monotonically $\sim 1/t$ (checkpoints at 25\%, 50\%, 75\%, 100\% of the run decrease throughout), the signature of linear separation growth compressed by renormalization, not of a positive-exponent plateau. The nominal values $3$--$5\times10^{-5}$ are consistent between the two configurations but, as the decisive diagnostic shows, overestimate the true exponent.

\begin{figure}[htbp]
\centering
\includegraphics[width=0.85\textwidth]{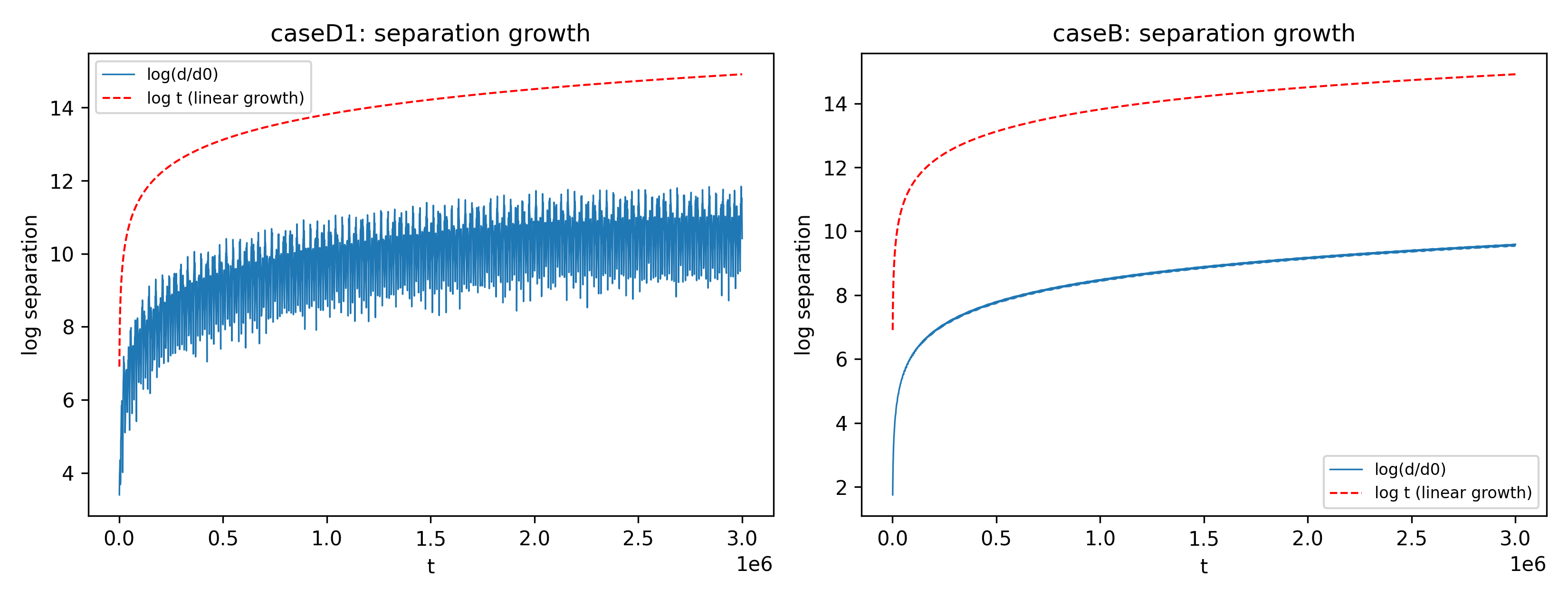}
\caption{Unrenormalized separation growth $d(t)$ (diagnostic (b), $d_0 = 10^{-9}$, $3\times10^{6}$ time units) for the five cases of Table~\ref{tab:diagnostics}, on doubly logarithmic axes with reference slopes. All four cases labeled chaotic in the example paper grow linearly in time (log--log slope $0.96$--$1.01$; exponential-fit slopes $\leq 4.6\times10^{-7}$), and Case D1 saturates. No case exhibits exponential separation.}
\label{fig:growth}
\end{figure}

\noindent\emph{(ii) The growth diagnostic is unambiguous.} Over $3\times10^{6}$ time units the inter-trajectory separation grows from $10^{-9}$ to only $3\times10^{-5}$--$6\times10^{-4}$. Fits of $\log d$ vs.\ $t$ give slopes $\leq 4.6\times10^{-7}$ in all four ``chaotic'' cases, i.e.\ an upper bound $\lambda_{\max} \lesssim 5\times10^{-7} \approx 0$; fits of $\log d$ vs.\ $\log t$ give slopes $0.96$--$1.01$, the textbook linear-in-time growth of regular orbits. Case D1 saturates outright (bounded oscillating separation), as expected for a near-resonant periodic/quasi-periodic orbit (Fig.~\ref{fig:growth}).

\begin{figure}[htbp]
\centering
\includegraphics[width=0.85\textwidth]{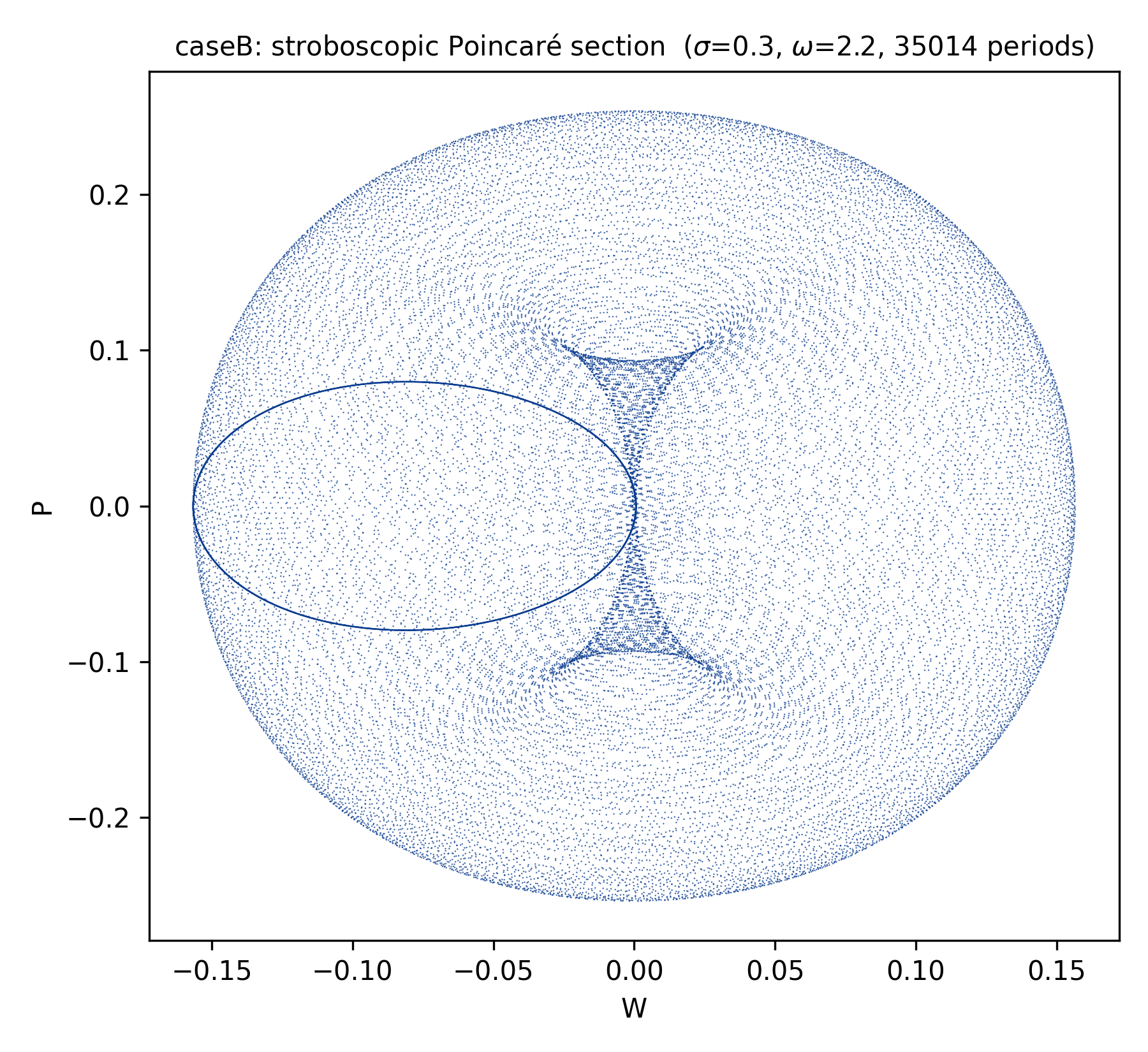}
\caption{Stroboscopic Poincar\'e section of Case B ($\sigma = 0.3$, $\omega = 2.2$, IC $(0,0.01)$; $\geq 3.5\times10^{4}$ forcing periods): a thin regular invariant band. The inset structure is resolved in Fig.~\ref{fig:zoomB}. This orbit was labeled chaotic in the example paper's Fig.~3.}
\label{fig:poincareB}
\end{figure}

\begin{figure}[htbp]
\centering
\includegraphics[width=0.85\textwidth]{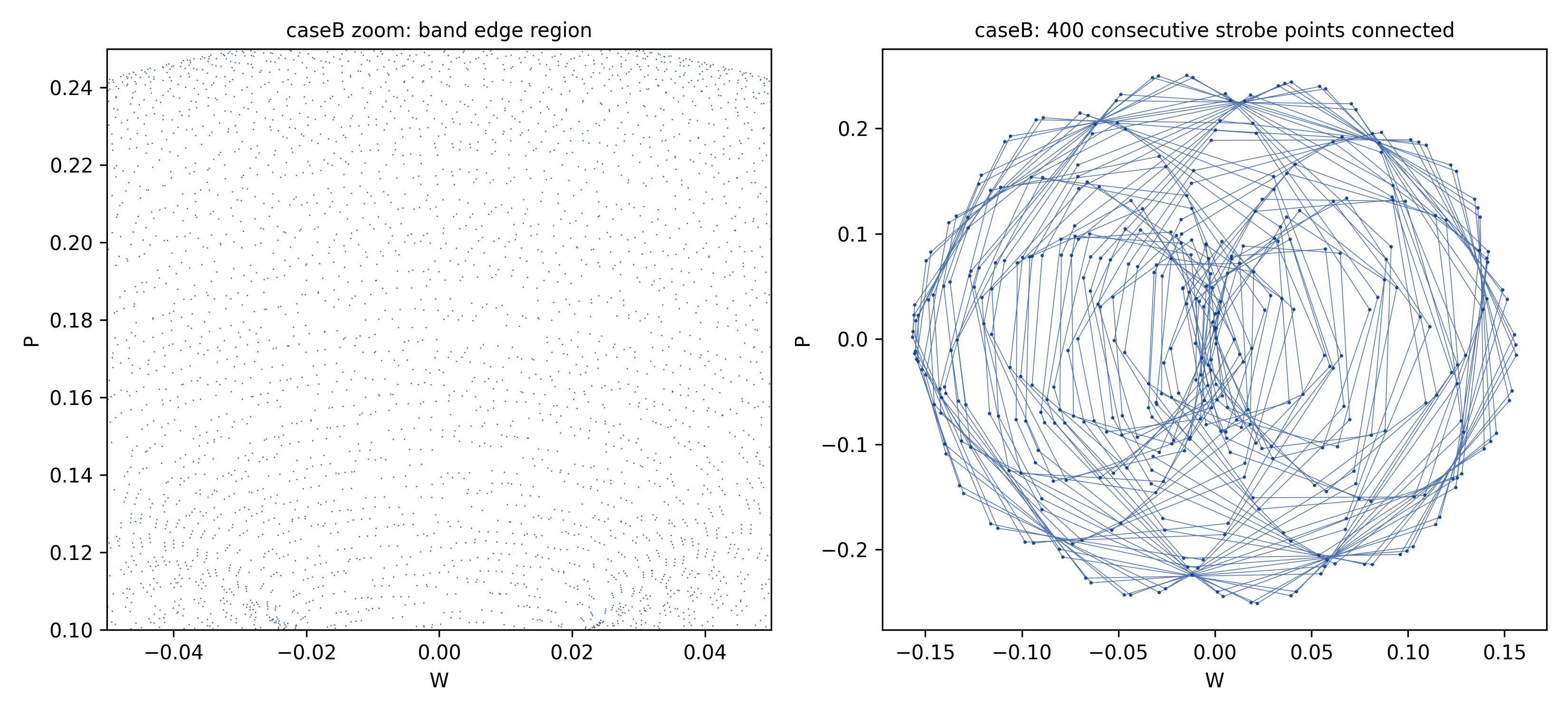}
\caption{Magnification of the Case B section: the band resolves into a smooth island chain --- the hallmark of a resonant quasi-periodic torus, not of a stochastic layer.}
\label{fig:zoomB}
\end{figure}

\begin{figure}[htbp]
\centering
\includegraphics[width=0.85\textwidth]{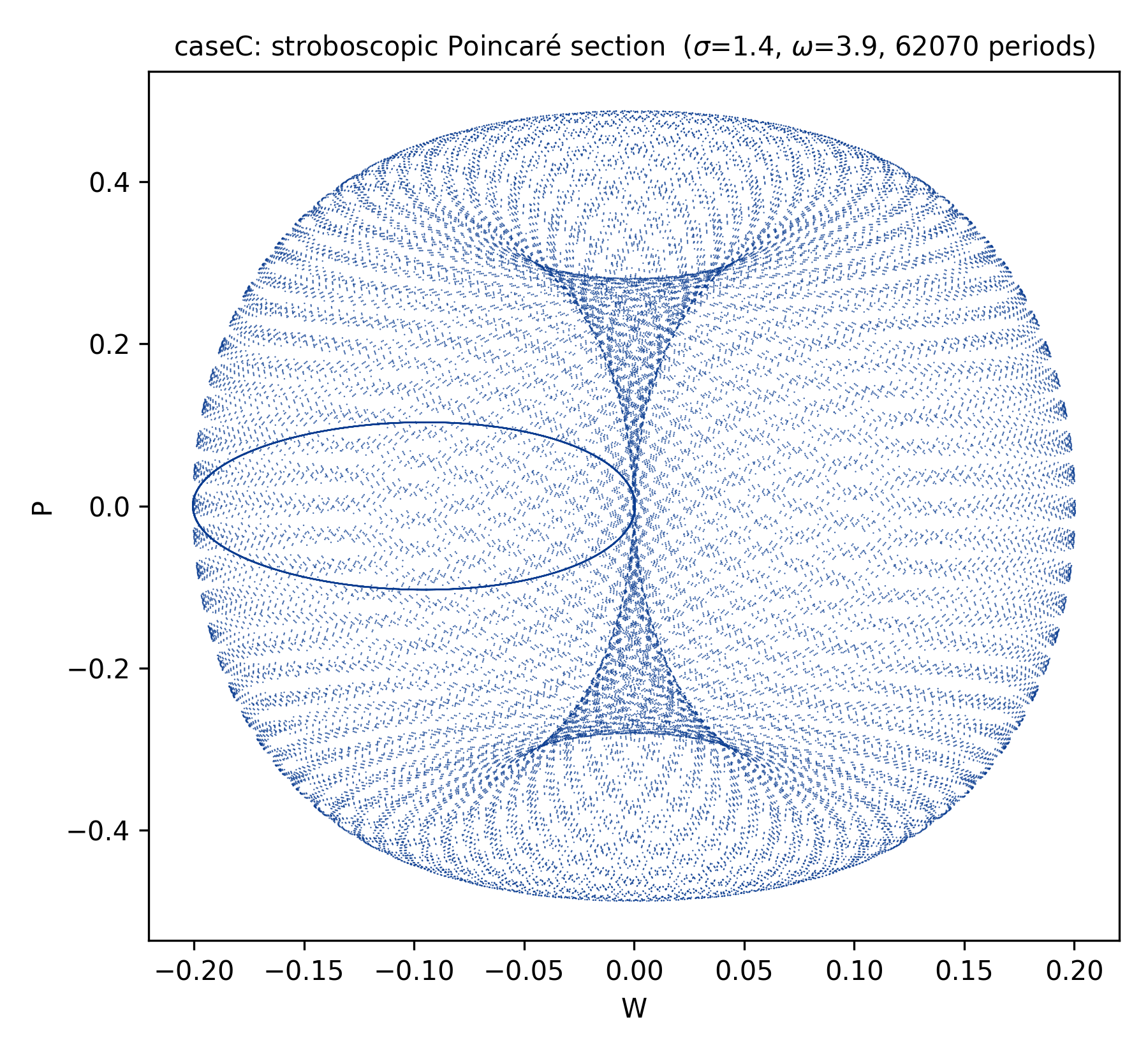}
\caption{Stroboscopic Poincar\'e section of Case C ($\sigma = 1.4$, $\omega = 3.9$, IC $(0,0.01)$): a thin closed invariant curve (2-torus). Labeled chaotic in the example paper's Fig.~4.}
\label{fig:poincareC}
\end{figure}

\begin{figure}[htbp]
\centering
\includegraphics[width=0.85\textwidth]{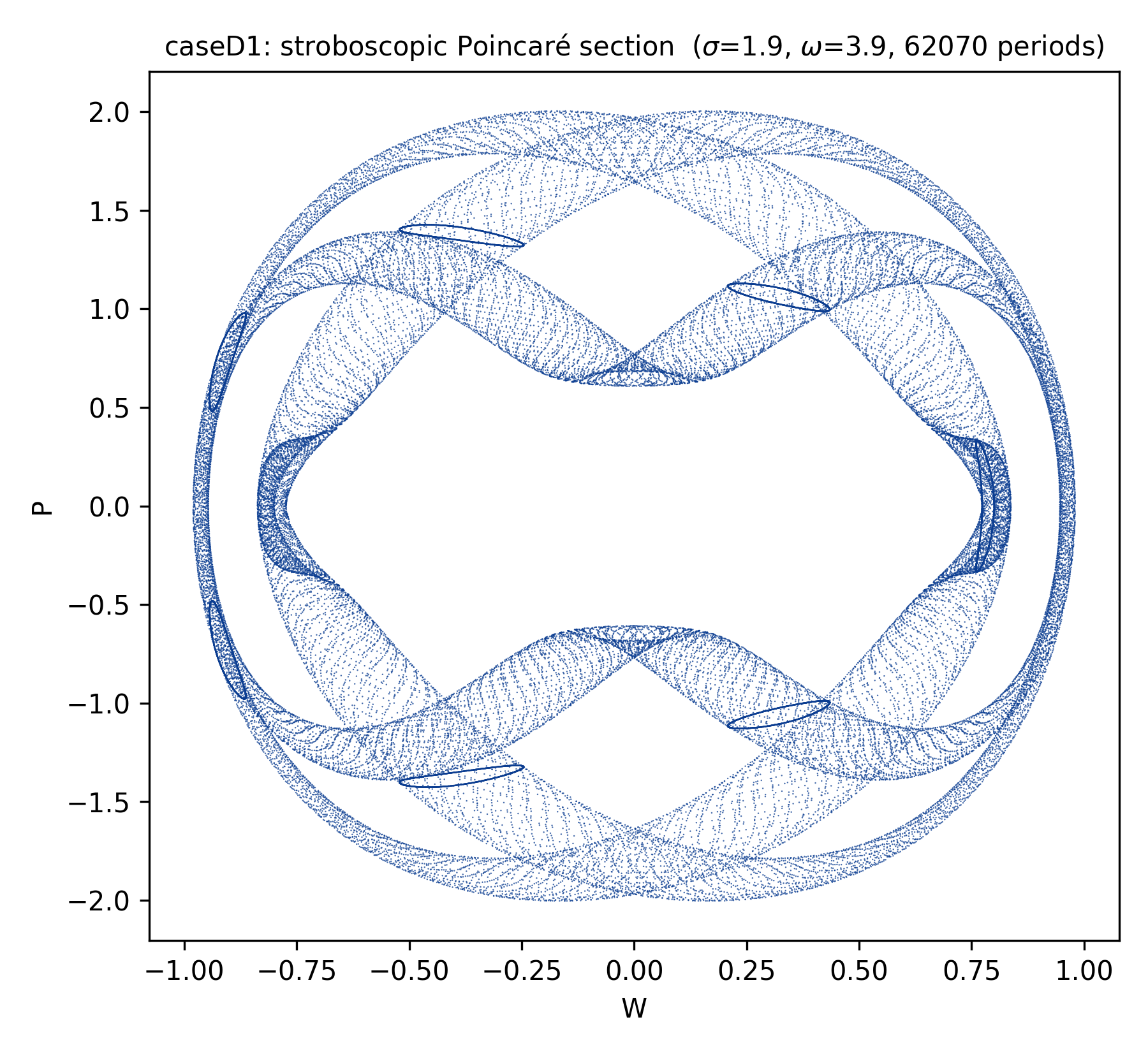}
\caption{Stroboscopic Poincar\'e section of Case D1 ($\sigma = 1.9$, $\omega = 3.9$, IC $(0.8,0)$): a closed invariant curve, consistent with the example paper's quasi-periodic label --- the one classification of its chaos section that our diagnostics confirm.}
\label{fig:poincareD1}
\end{figure}

\noindent\emph{(iii) The geometry confirms regularity.} All five Poincar\'e sections are smooth closed invariant curves (shown for Cases B, C and D1 in Figs.~\ref{fig:poincareB}--\ref{fig:poincareD1}; Cases D2 and D3 are similar and are omitted for brevity): Case B shows a thin regular band with an island-chain (resonance) structure, Case C a narrow closed 2-torus ring, and Cases D1--D3 simple closed curves. There is no area-filling scatter anywhere. The spectra of Cases B and C consist of discrete line clusters rather than broadband continua, a third independent line of evidence for regularity. All orbits remain bounded ($\max|W| \leq 0.98$) with no integrator anomalies.

\noindent\textbf{Correct statement.} For all tested parameter sets of the example paper's chaos section, the dynamics of system \eqref{eq:forced} is regular: Cases B, C, D2, D3 are quasi-periodic orbits on invariant tori (Case B showing a resonant island-chain structure), and Case D1 is quasi-periodic/near-periodic as originally stated. No chaotic orbit was found at any of the parameter sets examined. We do not claim that system \eqref{eq:forced} admits no chaos anywhere in parameter space --- thin stochastic layers may exist near resonances --- but the specific assertions of the example paper are unsupported.

\subsection{The morphology of stroboscopic projections: a likely source of the misclassification}
\label{sec:morphology}

\begin{figure}[htbp]
\centering
\includegraphics[width=0.85\textwidth]{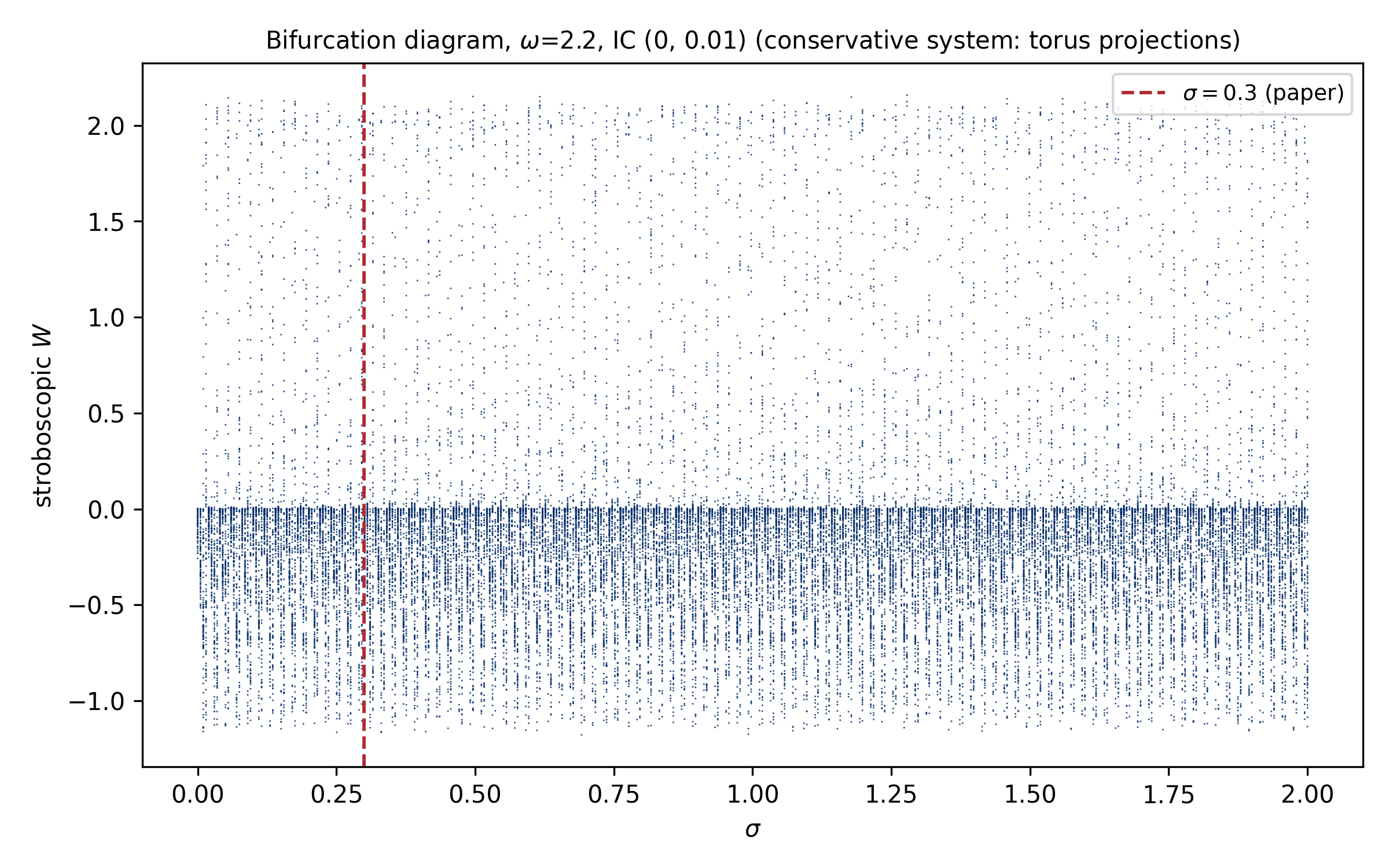}
\caption{Stroboscopic $\sigma$-sweep of system \eqref{eq:forced} at $\omega = 2.2$, IC $(0,0.01)$ (transient 300 forcing periods, 100 samples per $\sigma$; $\max|W| = 2.16$). The continuous broadband is the projection of invariant tori, not chaos; the vertical marker indicates the example paper's $\sigma = 0.3$ (Case B, proven regular in Table~\ref{tab:diagnostics}).}
\label{fig:bifw2p2}
\end{figure}

\begin{figure}[htbp]
\centering
\includegraphics[width=0.85\textwidth]{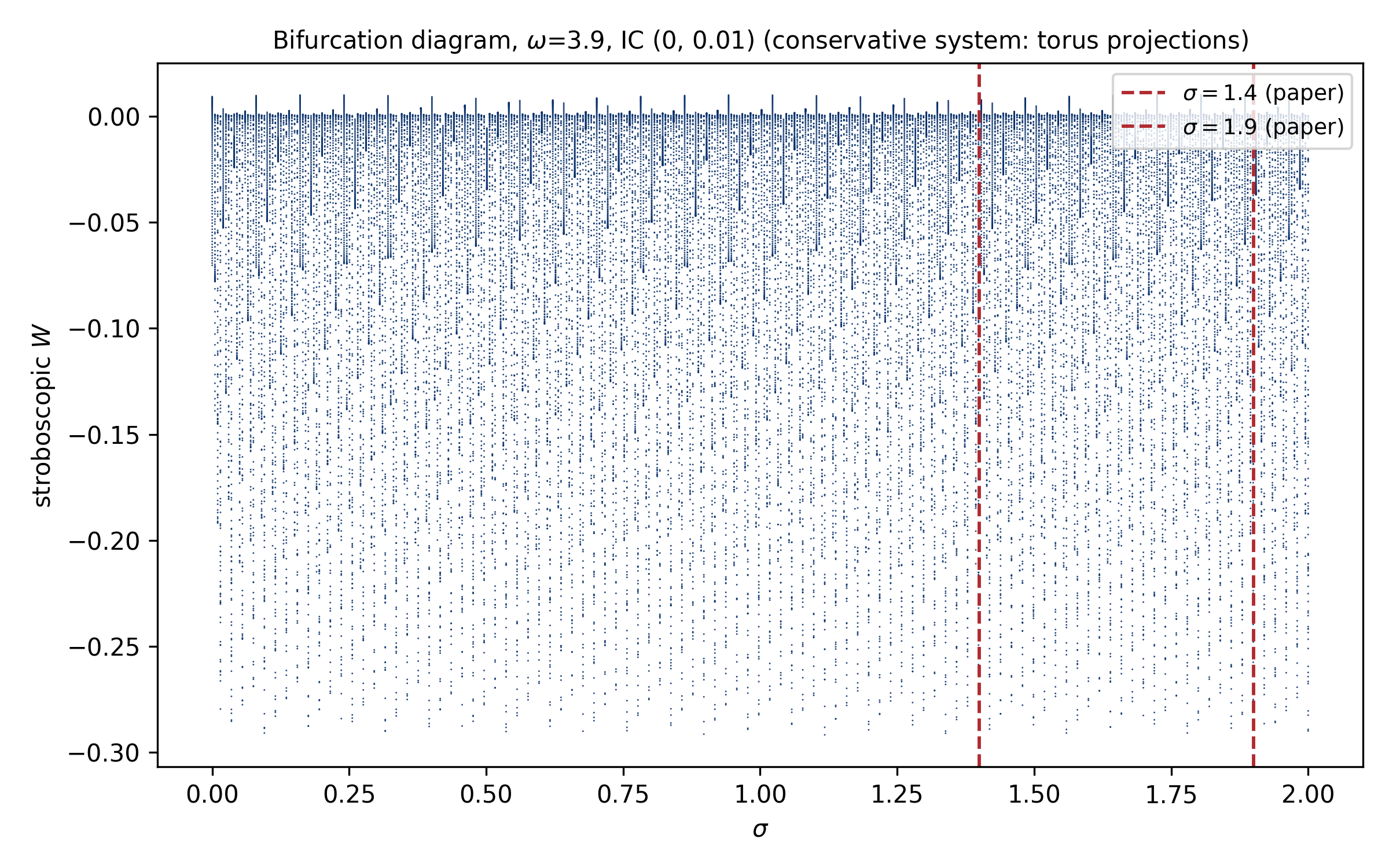}
\caption{As in Fig.~\ref{fig:bifw2p2} at $\omega = 3.9$ ($\max|W| = 0.29$); markers indicate the example paper's $\sigma = 1.4$ (Case C) and $\sigma = 1.9$ (Cases D1--D3), all proven regular in Table~\ref{tab:diagnostics}.}
\label{fig:bifw3p9}
\end{figure}

Why did the phase portraits look chaotic? Figures~\ref{fig:bifw2p2} and~\ref{fig:bifw3p9} show stroboscopic bifurcation diagrams of system \eqref{eq:forced} as a function of $\sigma \in [0, 2]$ (400 values integrated simultaneously; transient 300 forcing periods; 100 stroboscopic samples per value) for the two forcing frequencies used in the example paper. Both diagrams display broad, apparently structureless bands over essentially the whole parameter range --- superficially indistinguishable from the ``fully developed chaos'' region of a dissipative bifurcation diagram. The resemblance is, however, misleading: in a conservative system the stroboscopic projection of a quasi-periodic torus densely fills an interval (or a union of intervals), so \emph{every} regular orbit contributes a filled band. There is no period-doubling cascade, no periodic windows, no attractor. Diagrams of this kind ``look chaotic everywhere'' precisely because tori project onto bands. A phase portrait or a $\sigma$-sweep of this morphology cannot, by itself, distinguish quasi-periodicity from chaos; in the absence of damping-induced attractors, the visual complexity of a projection is not evidence of sensitive dependence. We regard this morphological trap as the most plausible origin of the example paper's misclassifications, and we expect it to operate throughout the pipeline literature, whose ``chaotic'' phase portraits are typically of exactly this type.

\subsection{A cautionary by-product: the 0--1 test fails on these orbits}
\label{sec:01test}

\begin{figure}[htbp]
\centering
\includegraphics[width=0.85\textwidth]{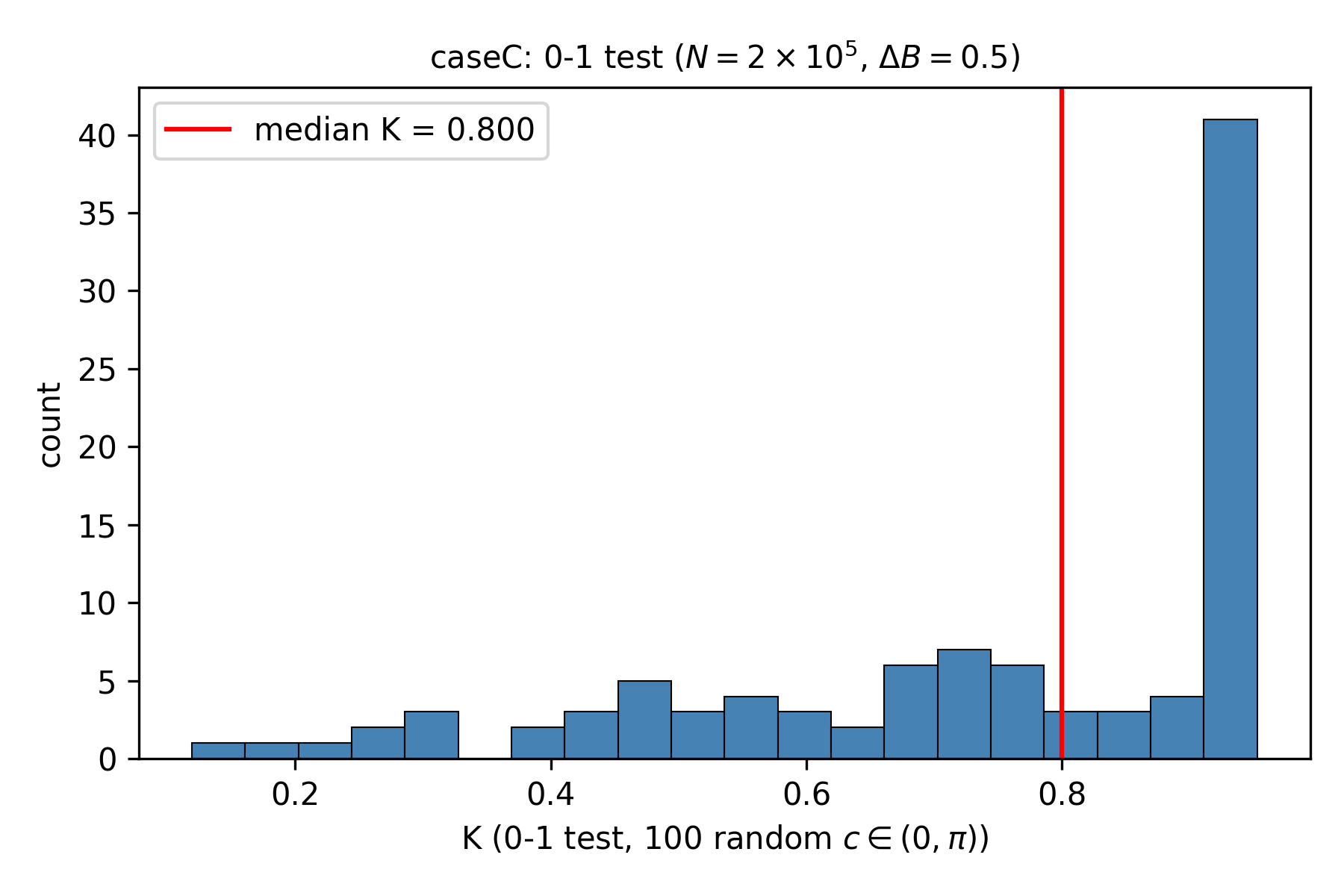}
\caption{Gottwald--Melbourne 0--1 test for Case C (a regular orbit, cf.\ Table~\ref{tab:diagnostics} and Fig.~\ref{fig:poincareC}): median $K = 0.800$ over 100 random $c \in (0,\pi)$ --- a false positive.}
\label{fig:01testC}
\end{figure}

Our cross-checking produced a methodological by-product worth reporting. The Gottwald--Melbourne 0--1 test returned a median $K = 0.800$ for Case C and $K = 0.681$ for Case D3 (Fig.~\ref{fig:01testC}) --- values that, taken at face value, would classify these orbits as chaotic, in direct contradiction with the Lyapunov bound ($\lambda_{\max} \lesssim 5\times10^{-7}$), the linear separation growth, the closed-curve Poincar\'e sections, and the discrete spectra. This is a false positive, not an implementation error: our implementation passes the standard calibration signals (periodic and quasi-periodic synthetic data give $K \approx 0$; the logistic map at $r = 3.9$ and white noise give $K \approx 0.998$), and the elevated $K$ values are stable under variation of $n_{\mathrm{cut}}$ ($N/20 \to N/5$) and of the random seed (Case C remains in $0.70$--$0.83$).

\begin{figure}[htbp]
\centering
\includegraphics[width=0.85\textwidth]{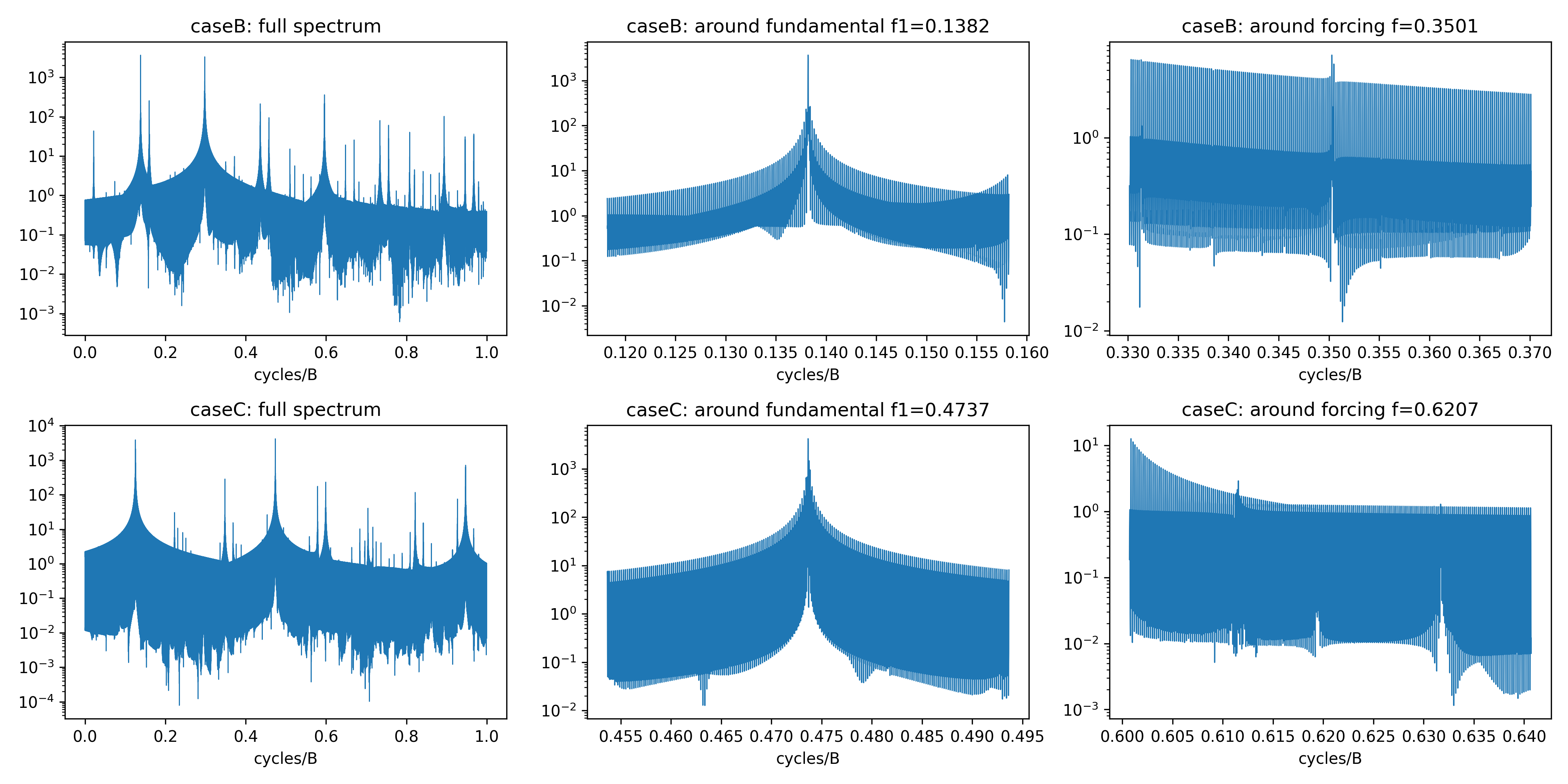}
\caption{High-resolution spectrum of the Case C time series around the fundamental peaks: dense clusters of 485--916 lines spaced $\sim 3\times10^{-5}$ cycles/$\mathcal{B}$, revealing a slow modulation (period $\gtrsim 3\times10^{4}$) that inflates the mean square displacement of the 0--1 test within any practical truncation window.}
\label{fig:spectral}
\end{figure}

The mechanism is visible in the spectrum (Fig.~\ref{fig:spectral}): around each fundamental peak of these orbits sit dense clusters of $485$--$916$ resolved spectral lines with spacing $\sim 3\times10^{-5}$ cycles/$\mathcal{B}$, i.e.\ the dynamics carries a slow modulation with period $\gtrsim 3\times10^{4}$ time units. Within any truncation window $n_{\mathrm{cut}} \ll 3\times10^{4}$, this slow component acts as a secular drift of the translation variables of the 0--1 test, producing a growing mean square displacement for almost every $c$ and hence an inflated $K$. In other words, the 0--1 test cannot distinguish a quasi-periodic orbit with a very slow third frequency from a weakly chaotic one unless the record length vastly exceeds the modulation period. The lesson generalizes: in undamped conservative systems, where KAM theory makes quasi-periodicity the generic outcome and chaos at most a thin-layer phenomenon, \textbf{no single statistic --- not visual phase portraits, not the 0--1 test, and not a single finite-time Lyapunov estimate --- is a reliable arbiter of chaos}. Convergent evidence from Lyapunov exponents with demonstrated $1/t$ decay, unrenormalized growth diagnostics, and Poincar\'e-section geometry should be required.

\section{Further remarks on the example paper}
\label{sec:remarks}

We collect four secondary observations, in decreasing order of significance. They are specific to the example paper, but each is symptomatic of recurring habits in the pipeline literature.

\noindent\textbf{(i) Conformable-derivative triviality.} The conformable derivative \cite{khalil2014} of a differentiable function satisfies $D_t^{\gamma} w = t^{1-\gamma}\, dw/dt$. Because the traveling-wave variable $\mathcal{B} = \frac{1}{\gamma}\left(l_1 x^{\gamma} + l_2 y^{\gamma} + l_3 z^{\gamma} - l_4 t^{\gamma}\right)$ is \emph{linear} in $t^{\gamma}/\gamma$ (and in $x^{\gamma}/\gamma$, etc.), the change of independent variables $\tau = t^{\gamma}/\gamma$, $\xi = x^{\gamma}/\gamma$, $\eta = y^{\gamma}/\gamma$, $\zeta = z^{\gamma}/\gamma$ converts the ``fractional'' equation (the example paper's Eq.~(7)) \emph{identically} into the integer-order second WBBM equation (its Eq.~(5)). Every dynamical and soliton result of the paper is therefore a result about the integer-order equation restated in rescaled coordinates; no memory or nonlocal effect --- the usual motivation for fractional modeling --- is present. This triviality of the conformable framework has been analyzed in detail by Abdelhakim and Machado \cite{abdelhakim2019}; we note it because the example paper's framing attributes physical significance to the fractional order $\gamma$.

\noindent\textbf{(ii) Parameter table inconsistency.} In Case 2 of the bifurcation section, the parameter list reads ``$l_1 = l_2 = l_3 = 1$, and $l_2 = 2$'': $l_2$ appears twice with contradictory values. Comparison with the stated regime $\alpha > 0$, $m < 0$ (and with the numerical values $\alpha = 1/2$, $m = -1/2$ used in the example paper's Fig.~1b) shows that the intended assignment is $l_4 = 2$.

\noindent\textbf{(iii) Non-standard terminology.} The term ``hyperperiodic orbits'' used for the large closed curves encircling both wells in Case 2 is not standard in the dynamical-systems literature; these are simply periodic orbits of the outer family (encircling the figure-eight separatrix).

\noindent\textbf{(iv) Novelty claims.} The solutions $w_1$--$w_5$ (Jacobi $\mathrm{sn}$, $\mathrm{dn}$, $\mathrm{cn}$ functions and their $\tanh$/$\operatorname{sech}$ limits) are the classical, long-known exact solutions of the $\phi^4$-type potential \eqref{eq:hamiltonian}; their derivation via the planar-dynamics method of Liu and Li \cite{liu2002} is standard. The novelty asserted in the example paper's Section~10 should therefore be read as pertaining to the application, not to the solutions themselves --- and the dynamical classifications that constitute its genuinely new claims are precisely those that do not withstand quantitative scrutiny.

\section{Recommendations for chaos claims in this literature}
\label{sec:recommendations}

The case study above is one paper, but the failure modes it exhibits are structural features of the pipeline genre. We therefore distill our analysis into a checklist that authors, referees, and editors can apply to any paper of this type --- traveling-wave reduction followed by forced-oscillator chaos claims.

\noindent\textbf{(a) Report the largest Lyapunov exponent, with convergence evidence.} A chaos claim should be accompanied by a finite-time largest Lyapunov exponent computed by a stated algorithm (e.g.\ Benettin's two-trajectory method \cite{benettin1980}), with stated initial separation, renormalization interval, integration length, and --- critically --- a convergence diagnostic. A running estimate that decays $\sim 1/t$ indicates a regular orbit, whatever its instantaneous value; only a plateau bounded away from zero supports chaos. A bare ``Lyapunov plot'' without methodological detail, as in the example paper's Fig.~5c, carries no evidentiary weight.

\noindent\textbf{(b) Show stroboscopic Poincar\'e sections.} A section recorded once per forcing period over $\gtrsim 10^{4}$ periods immediately separates invariant closed curves (quasi-periodicity) from area-filling stochastic layers (chaos). This single figure would have sufficed to prevent all four erroneous classifications in the example paper.

\noindent\textbf{(c) Discuss the conservative structure explicitly.} If the reduced system is undamped --- as traveling-wave reductions of conservative PDEs almost always are --- the paper should acknowledge the KAM framework \cite{lichtenberg1992}: quasi-periodic tori are the generic outcome, chaos is at most a thin-layer phenomenon, and the claim of chaos at a \emph{specific} parameter set requires locating that set inside a stochastic layer, ideally with an estimate of the layer's extent. In single-well reductions without a separatrix, the prior against chaos is especially strong.

\noindent\textbf{(d) Require at least two independent diagnostics.} No single statistic is decisive in conservative systems. We have shown that the 0--1 test --- calibrated and correctly implemented --- returns $K \approx 0.80$ (a ``chaotic'' verdict) for an orbit that is provably a torus, because slow modulations with dense spectral line clusters inflate its mean square displacement. Finite-time Lyapunov estimates can likewise produce small positive values as renormalization artifacts. Convergent evidence --- e.g.\ a Lyapunov exponent with demonstrated convergence \emph{plus} a Poincar\'e section, or a growth diagnostic \emph{plus} a spectrum --- should be the minimum standard.

\noindent\textbf{(e) Do not read bifurcation diagrams morphologically.} In dissipative systems, broadband regions of a parameter sweep correlate with chaos because attractors alternate with periodic windows. In conservative systems the stroboscopic projection of \emph{every} torus fills a band, so sweeps look chaotic everywhere (our Figs.~\ref{fig:bifw2p2} and~\ref{fig:bifw3p9}, computed entirely from regular orbits, are indistinguishable in kind from published ``chaotic'' sweeps). A broadband bifurcation diagram in an undamped system is evidence of nothing except the presence of tori.

\noindent\textbf{(f) Editorial standards.} Journals publishing this genre would improve the literature materially by requiring quantitative diagnostics (per (a)--(d)) for any claim of chaos or quasi-periodicity, in the spirit of Sprott's publication standard \cite{sprott2011}. The marginal cost is small: for a two-dimensional forced oscillator, a converged Lyapunov exponent and a Poincar\'e section require minutes of computation. The example paper was received on 4 June 2024 and accepted on 8 July 2024; a one-month review cycle leaves little room for checking dynamical claims, which makes explicit checklists all the more valuable at the refereeing stage.

\section{Conclusions}
\label{sec:conclusions}

We have argued that a popular pipeline in the fractional nonlinear-wave literature --- traveling-wave reduction, planar Hamiltonian analysis, periodic forcing, and visually assigned dynamical labels --- systematically overclaims chaos, and we have demonstrated the failure modes in detail on a recent representative example, the second fractional WBBM study of Ullah, Ali and Roshid \cite{ullah2024}. For that example we established: (i) the linear stability conclusion is analytically untenable --- the dispersion relation is real for all real wave numbers, every Fourier mode is neutrally stable, and the reported ``unstable propagation'' is a pole (resonant surface) misidentified as a temporal instability; (ii) the unforced regime labeled ``quasi-periodic'' is necessarily periodic, both by Hamiltonian structure and by spectral analysis at energy conservation $2.7\times10^{-15}$; (iii) all four chaos assertions fail every quantitative diagnostic --- largest Lyapunov exponents bounded by $5\times10^{-7}$, separation growth linear in time (log--log slopes $0.96$--$1.01$) over $3\times10^{6}$ time units, smooth closed invariant curves in stroboscopic Poincar\'e sections, discrete line spectra --- while the single quasi-periodic assertion is confirmed; and (iv) the equilibrium classification and phase portraits are correct.

The general lessons extend beyond the example. In undamped, near-integrable forced oscillators, KAM theory makes regular tori the generic outcome and chaos a thin-layer phenomenon that must be proved, not presumed. Phase-portrait morphology and broadband parameter sweeps cannot perform this proof --- conservative tori project onto bands that mimic chaos. Single statistics can actively mislead: the 0--1 test produced a false positive ($K \approx 0.80$) on a provably regular orbit through a slow-modulation mechanism that is intrinsic to this class of systems. The checklist of Section~\ref{sec:recommendations} --- converged Lyapunov exponents, stroboscopic Poincar\'e sections, explicit engagement with the conservative structure, and cross-validation by at least two independent diagnostics --- is inexpensive to apply and would have prevented every error documented here. We hope it proves useful to authors, referees, and editors working in this literature.

\appendix

\section{Numerical methods}
\label{app:methods}

All computations were performed with custom scripts; intermediate data and figure scripts are available from the authors upon request.

\noindent\textbf{High-accuracy orbit integration.} DOP853 (8th-order Dormand--Prince) as implemented in SciPy, with $\mathrm{rtol} = 10^{-11}$, $\mathrm{atol} = 10^{-13}$ for production records ($\mathrm{rtol} = 10^{-12}$, $\mathrm{atol} = 10^{-14}$ for the $\sigma = 0$ energy-conservation tests). Transients of $10^{4}$ time units were discarded; sampled records span $10^{5}$ time units at $d\mathcal{B} = 0.5$ ($N = 2\times10^{5}$ samples) plus $\geq 3.5\times10^{4}$ stroboscopic points at the forcing period $2\pi/\omega$.

\noindent\textbf{Benettin largest Lyapunov exponent.} Two-trajectory method \cite{benettin1980} with a fixed-step RK4 integrator (JIT-compiled), initial separation $d_0 = 10^{-9}$. Main configuration: renormalization interval $\Delta\mathcal{B} = 1$, step $10^{-3}$; robustness configuration: $\Delta\mathcal{B} = 5$, step $5\times10^{-4}$. Total length $2\times10^{5}$ after a transient of $10^{4}$; reported uncertainties are standard errors of the mean over 25 blocks. Convergence was assessed from running estimates at the 25\%, 50\%, 75\% and 100\% checkpoints; all runs show monotone $\sim 1/t$ decay.

\noindent\textbf{Unrenormalized growth diagnostic.} Pairs of trajectories with $d_0 = 10^{-9}$, fixed-step RK4 with step $10^{-3}$, integrated for $3\times10^{6}$ time units. Exponential fits of $\log d$ vs.\ $t$ and logarithmic fits of $\log d$ vs.\ $\log t$ were performed over the full record; separations remain $\leq 6\times10^{-4}$ throughout, far below the attractor size, so no saturation contaminates the fits.

\noindent\textbf{Spectral analysis.} Hann-windowed FFT of the detransiented records; peaks with amplitude $\geq 10^{-6}$ of the dominant peak retained. Dense-cluster counts (485--916 lines) were obtained from high-resolution spectra with line spacing $\sim 3\times10^{-5}$ cycles/$\mathcal{B}$.

\noindent\textbf{0--1 test.} Gottwald--Melbourne test \cite{gottwald2004} with $N = 2\times10^{5}$ samples at $d\mathcal{B} = 0.5$, 100 random $c \in (0, \pi)$, mean square displacement computed by the exact FFT method with the oscillatory correction term removed, $n_{\mathrm{cut}} = N/10$ (sensitivity checked for $N/20 \le n_{\mathrm{cut}} \le N/5$). Calibration: synthetic periodic and quasi-periodic signals $\to K \approx 0$; logistic map at $r = 3.9$ and white noise $\to K \approx 0.998$.

\noindent\textbf{$\sigma$-sweeps.} 400 values of $\sigma \in [0,2]$ integrated simultaneously (vectorized 800-dimensional system), DOP853 with $\mathrm{rtol} = 10^{-9}$, $\mathrm{atol} = 10^{-11}$, IC $(0, 0.01)$; after a transient of 300 forcing periods, 100 stroboscopic $W$-values recorded per $\sigma$.

\noindent\textbf{Code availability.} The scripts used for all computations and figures are available from the authors upon request.

\end{document}